\documentclass[aps,epsfig,prb,twocolumn,showpacs,floatfix]{revtex4}

\usepackage{graphicx}
\usepackage{dcolumn}
\usepackage{epsfig}
\usepackage{float}
\usepackage{bm}

\usepackage{bibentry}

\begin{document}

\bibliographystyle{apsrev}

\title{Infrared optical    absorption spectra of CuO single crystals: Fermion-spinon band and    dimensional crossover of the antiferromagnetic order}

\author{SeongHoon Jung}
\author{Jooyeon Kim}
\author{E. J. Choi}
\affiliation{Department of Physics, University of Seoul, Seoul
130-743, Republic of Korea}

\author{Y. Sekio}
\author{T. Kimura}
\affiliation{Division of Materials Physics, Graduate School of
Engineering Science, Toyonaka, Osaka University, Osaka 560-8531,
Japan}

\author{J. Lorenzana}
\affiliation{SMC-INFM-CNR, ISC-CNR and Dipartimento di Fisica,
Universit\`a di Roma ``La Sapienza'', P. Aldo Moro 2, 00185 Roma, Italy}

%
%

\begin{abstract}
We have obtained mid-infrared optical absorption spectra of the
$S=1/2$ quasi one-dimensional CuO using polarized  transmission
measurement and interpreted the spectra in terms of phonon assisted magnetic
excitations. When the electric field is parallel to the main
antiferromagnetic direction a $\Delta$ shaped peak is observed
with the maximum at $\omega = 0.23$~eV which is attributed to spinons
along Cu-O chains. At low temperatures in the antiferromagnetic phase
another peak appears at $\omega = 0.16$~eV which is
attributed to two-magnon absorption but the spinon peak remains. This behavior is interpreted as
due to a dimensional crossover where the low temperature three-dimensional
magnetic phase keeps short range characteristics of a one-dimensional magnet.
\end{abstract}

\pacs{74.72.-h,71.10.Pm,74.25.Gz, 78.20.-e}

\maketitle

Properties of one dimensional systems are radically different from
those of higher dimension. For Heisenberg one-dimensional (1D)
antiferromagnets (AF) the ground state is a spin liquid with fermionic
excitations called spinons rather than the usual, propagating spin waves,
of ordered higher dimensional systems.
A fundamental question in condensed matter physics is what of these anomalous
properties and to which extent can survive in anisotropic but higher
dimensional systems.  For a quasi-1D system formed by chains bridged
by small perpendicular interactions the ground state is expected to be
ordered, however a spin liquid may still be a good approximation for
the ground state. Furthermore 2D quantum antiferromagnets, although
ordered at $T=0$, are believed to be close to a spin liquid ground
state. This has been widely studied in connection with high
temperature superconductors as it has been proposed that
superconductivity emerges
from a spin liquid\cite{and04}. Even in an ordered state, intermediate
to short range spin liquid
characteristics may reflect on the response at finite frequencies,
as proposed by Anderson and collaborators to explain
puzzling side bands in optical spectra of 2D AF
cuprates\cite{cho01}.

In this communication we study the mid-infrared spectrum of
CuO (tenorite) due to magnetic excitations infrared (IR) active by
the Lorenzana and Sawatzky  (LS)
 phonon-assisted mechanism\cite{lor95,lor95b}.
CuO  is a quasi-1D spin-1/2 compound (Cu$^{2+}$) which becomes a 3D
antiferromagnet\cite{yan89b,for89} at low temperatures $T<T_{N1}=
210$K. It has an intermediate phase at $T_{N1}<T<T_{N2}=230$K which
is a spiral and has attracted recent attention for being a
high-$T_c$ multiferroic\cite{kim08}. We show that the mid-IR
spectrum of CuO at $T>T_{N2}$ can be interpreted as due to spinon
excitations as found for Sr$_2$CuO$_3$, the well known 1D S=1/2 AF material \cite{suz96,lor97}. In the ordered phase ($T<T_{N2}$) a feature
due to the propagation of two spin waves appears similar to the
bimagnon peak of 2D cuprates\cite{lor95,lor95b}. The high energy
side bands however remain strong and evolve continuously from the
fermionic spectrum in strong analogy with the scenario of
Ref.~\cite{cho01} in which fermionic excitations persist in an
ordered system.

%
%
\begin{figure}[tb]
\begin{center} 
\includegraphics[width=11cm,bb=100 100 800 580]{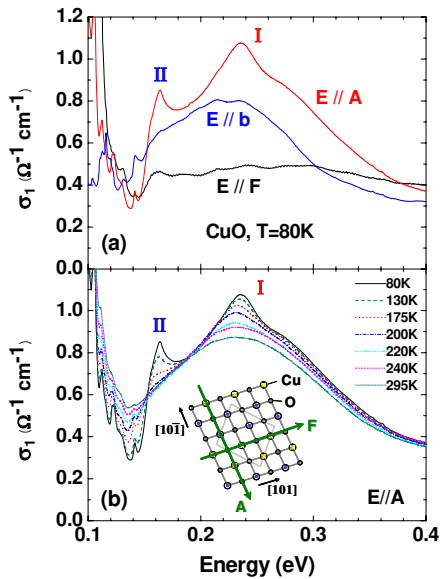}
\end{center}
\caption{(Color online) (a) Mid-infrared conductivity of CuO crystal when IR
electric field is polarized along the A-axis  ({\bf E}//A), F-axis
({\bf E}//F) and $b$-axis ({\bf E}//$b$) respectively.
(b)Temperature dependent optical conductivity $\sigma_{1}(\omega)$
for {\bf E}//A mode. The inset shows the spin ordering at T$<$
T$_{N_{1}}$ in the $ac$-plane. Here A and F represent the
antiferromagnetic (A) and ferromagnetic (F) axes, respectively.
}\label{fig:fig1}
\end{figure}

The tenorite CuO single crystal with the monoclinic lattice
structure was grown using the floating zone method \cite{kim08}. At
$T < T_{N_{1}}$ the Cu$^{2+}$ spins (S=1/2) are parallel to the
[010] direction (=$b$-axis) pointing normal to the $ac$-plane (see
inset of Fig.~\ref{fig:fig1} (b)). Within the plane they show antiferromagnetic
alignment along the [10$\bar{1}$] direction and ferromagnetic order
along the [101] direction. We label the two magnetic directions as
A=[10$\bar{1}$] and F=[101] respectively. To probe IR response of
the three principle axes A,$b$, and F, we have prepared three
samples with (A,F),(A,$b$) and ($b$,F) faces respectively with
typical area of 5mm $\times$ 5mm. For IR transmission measurement,
they were mechanically polished to $\sim$50 $\mu$m thickness.
Incident IR radiation  was linearly polarized such that the electric
field $\bf E$ was parallel to the magnetic axes.  For example in the
(A,F) sample the transmission $T(\omega)$ was measured with the IR
electric field $\bf E$ and magnetic field $\bf B$ set as ($\bf
E$//A, $\bf B$//F) and ($\bf E$//F, $\bf B$//A). With the three
samples we collect the full set of data, ($\bf E$, $\bf B$)// (A,F),
(A,$b$),(F,$b$),(F,A),($b$,F) and ($b$,A). From them we find that
$T(\omega)$ is determined by $\bf E$ but not dependent on $\bf B$.
For example, $T(\omega)$'s for ($\bf E$, $\bf B$)// (A,F) and
(A,$b$) were identical. It holds when $\bf E$ is polarized along C
and $b$ axes as well. In the monoclinic CuO, A and F are not
orthogonal but form an angle of 99.54$^{o}$. Therefore it is not
possible to get a ``pure'' F or A response but there is ``leakage''
in $T(\omega)$ although the effect is small. The optical
conductivity $\sigma_{1}(\omega)$ for the three modes $\bf E$//A,
$\bf E$//F, and $\bf E$//$b$ were extracted from $T(\omega)$ using
the standard Fresnel formula analysis.

Fig.~\ref{fig:fig1}(a) shows the optical conductivity of CuO taken
at $T=80$K when {\bf E} is parallel to A, F, and $b$ axes. In $\bf
E$//A mode, the mid-IR absorption consists of a broad $\Delta$
shaped peak labeled II with a much narrow peak superposed on the low
energy part labeled I.  These Peak I and II are absent in the
ferromagnetic axis $\bf E$//F. This is consistent with an electric
dipole operator\cite{lor95} containing a term ${\bf S}_i. {\bf S}_j$
which excites AF aligned spins in sites $i$,$j$.  In the $\bf E$//b
polarization we find an absorption bands with an intermediate
strength. The rapid rise of $\sigma_{1}(\omega)$ at $\omega <
140$~meV is due to the optical phonons. The infrared phonons of CuO
crystal were studied by Kuzmenko {\it et al.} from reflectivity
measurement \cite{kuz01} according to which the highest energy
phonon peak lies at 65~meV for $\bf E$//A. We have checked that the
tail of such phonon overlaps the rise of $\sigma_{1}(\omega)$ below
140~meV. The strength of the phonon at its center  frequency is
about 800$(\Omega\cdot$ cm)$^{-1}$, much greater than that of the
mid-IR peak. It shows that although both features are
\textbf{E}-active, the mid-IR absorption has a different activation
origin from the phonon.

According to the LS mechanism magnetic excitations which are not
directly dipole active, become visible in the IR spectra thanks to
the assistance of an optical phonon. At lowest order two magnons (or
two spinons in 1D) plus a phonon are created by the absorbed photon.
Generally the spectrum is similar to the magnetic Raman spectrum
with the difference that because the phonon carries momentum,
 an extra integration over the total momentum of the magnon
(or spinon) pair, $q$ is required, instead of the pair having zero
total momentum as in
Raman. Thus the spectrum may have sharp structures due to Van Hove
singularities which provide precious information on the excitations in
the system. Another important difference with Raman is that the
spectrum at $T=0$ is shifted by the phonon frequency thus appearing as
a phonon side band. The relevant phonon is usually a stretching mode
phonon whose frequency can be estimated from the highest optical
phonon. In the following we take $\omega_{ph}=65$~meV\cite{kuz01} for
the {\bf E}//A mode which we analyze in detail. The AF chains can also
contribute in the $b$ direction involving a $b$ polarized buckling
phonon which can add to interchain magnetic contributions involving
other phonons. Thus the {\bf E}//$b$ line shape has a mixed intrachain
-interchain character. This more complicated case probably relevant
for the multiferroic effect will be analyzed elsewhere.
%
%
\begin{figure}[tb]
$$\includegraphics[height=5cm,bb=0 30 300 230]{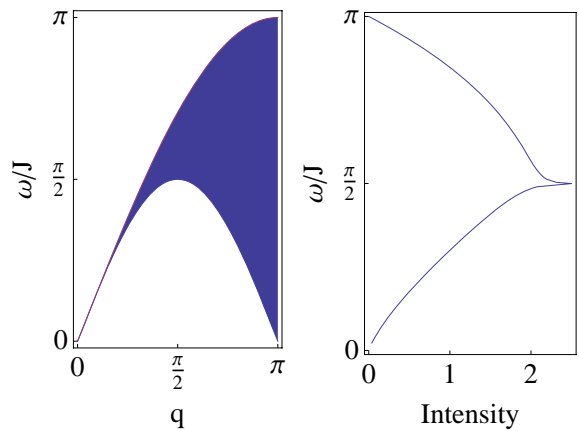}$$
\caption{(Color online) The left panel shows the spinon continuum (shaded region)
in the $\omega$-$q$ plane. The right panel shows  the line
  shape that results after momentum integration (neglecting the phonon).}\label{fig:fig2}
\end{figure}

We note that peak I, in {\bf E}//A mode, is similar to the two-magnon
peak observed in 2D
cuprates at $\sim 0.4 eV$ which is dominated by
bimagnons\cite{per93,per98,lor95,lor95b} while peak II  resembles the
$\Delta$ shaped peak observed in 1D cuprates\cite{suz96} and well
understood in terms of spinon absorption\cite{lor97}.
In a purely 1D AF Heisenberg system with exchange interaction $J$,
the dipole operator connects the
ground state with the two spinon continuum which is
limited by the des Cloizeaux-Pearson dispersion
relation\cite{clo62} (the shaded region in the left panel of Fig.~\ref{fig:fig2}). At $T=0$ the spectrum of spinon excitations can
be computed virtually exactly\cite{lor97} while at finite temperatures numerical
 and analytical results are available\cite{gag00}.  The lower edge of
 the continuum  has a saddle at momentum $q=\pi/2$ (setting the
 lattice spacing $a=1$) with energy
$\omega(q=\pi/2)=\pi J/2$.  When
integrated over momentum at $T=0$ a  Van Hove
singularity appears at $\omega=\omega_{ph}+\pi J/2$ corresponding
to the maximum of the absorption\cite{lor97} (c.f. Fig.~\ref{fig:fig2},
right panel).
Increasing the temperature the Van Hove
singularity broadens~\cite{gag00} but the maximum remains close by.
%
%
\begin{figure}[tb]
$$\includegraphics[height=6cm,bb=100 80 600 400]{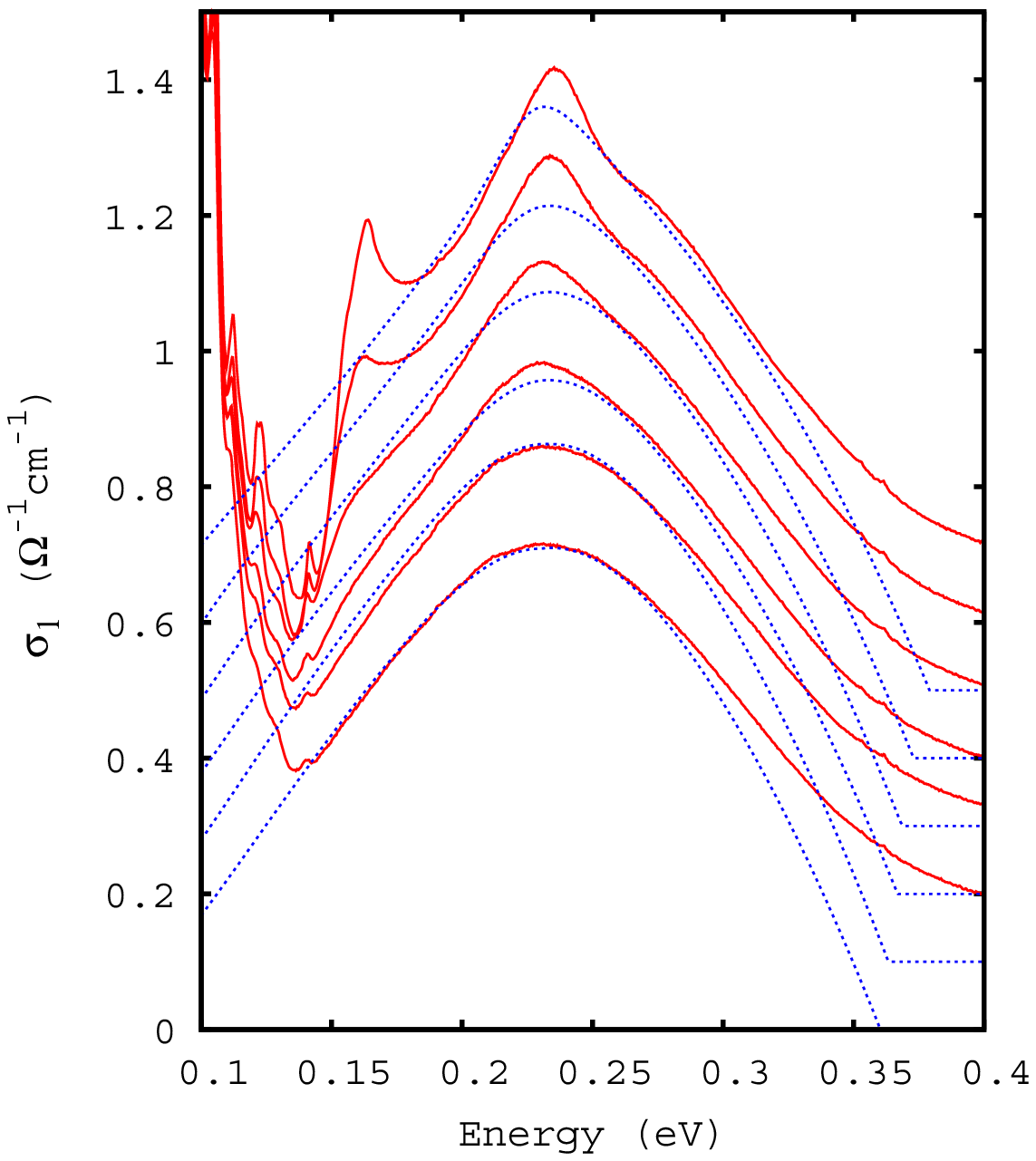}$$
\caption{(Color online) Full lines: Temperature dependent changes of experimental
  (full curves) and theoretical (dashed lines) $\sigma_{1}(\omega)$
of the {\bf E}//A polarization for (from top to bottom) $T=80$K, 150K,
200K, 220K, 260K and 295K. Curves where shifted by
$0.1\Omega^{-1}$cm$^{-1}$ for clarity.
A common constant background was subtracted to the experimental data.
For the theory we used an effective $J$ varying linearly with
temperature with $J(295K)=94$~meV and  $J(80K)=100$~meV and with
similar small adjustments to the intensity scale.}
\label{fig:fig3}
\end{figure}

Fig.~\ref{fig:fig3} compares the experimental data with the two spinon
theoretical ansatz of Ref.~\cite{lor97,gag00} for a purely 1D system
at different temperatures.  Surprisingly despite the substantial 3D
character of the magnetism in CuO, the main feature is in good
agreement with the theory with a weakly varying temperature dependent
effective $J$. At high temperatures the main difference is the high
energy tail which may be attributed to higher multispinon or
multiphonon/multispinon processes.
Thus above the ordering temperatures the
line shape is well explained by the fermionic spinon
excitations of a pure 1D system. Indeed at high
enough temperature spins on different chains are uncorrelated and the small
interchain interactions average to zero. As the temperature is lowered
the effect of the interchain interactions becomes to be relevant. One
effect is to renormalize upwards the effective $J$ along the chains due
to indirect exchange processes. This explains the weak temperature
dependence of $J$ needed to fit the experiment. Another effect will be
to drive 3D magnetic order as discussed below.
%
%
\begin{figure}[ht]
$$\includegraphics[height=4cm,bb=200 180 600 490]{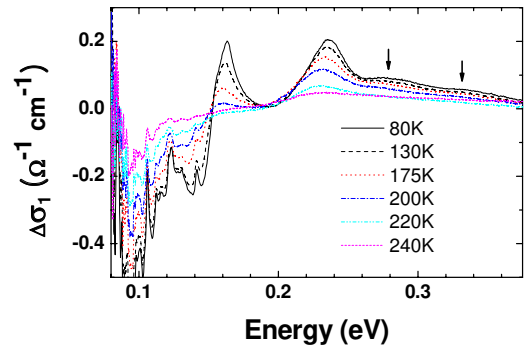}$$
\caption{(Color online) Optical conductivity difference $\Delta\sigma_{1}(T)\equiv
\sigma_{1}(T)-\sigma_{1}(295K)$.}
\label{fig:fig4}
\end{figure}

\begin{figure}[tb]
$$\includegraphics[height=6cm,bb=80 120 746 550]{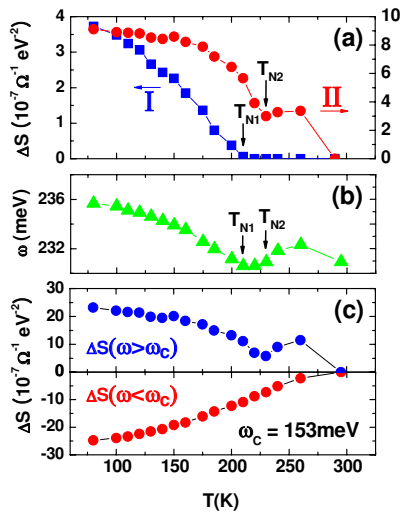}$$
\caption{(Color online) (a) Optical spectral weight change of the peak I and peak
II estimated from $\Delta S(T)= \int \Delta\sigma_{1}(T,
\omega)d\omega$. (b) Temperature dependence  of the maximum of peak
II.  (c) $\Delta S(T)$ for the low frequency region 80meV $<$ $\omega < $
153meV and for the high frequency mid-IR peaks at 153meV $<$ $\omega < $
375meV.}
\label{fig:fig5}
\end{figure}

In order to understand the origin of peak I we plot in
Fig.~\ref{fig:fig4} the change of
optical conductivity with respect to the room-$T$ data.
 Note that peak I and peak II show  different $T$-dependences; while
 peak II evolves gradually in this temperature range, consistently with
 the fact that $K_BT<<J$, peak I develops below $T_{N1}$ trough a
 reorganization of the low energy part of the spinon continuum.
This can be seen more clearly in Fig.~\ref{fig:fig5}(a) where we plot
the change in oscillator strength  of the two peaks as
 $\Delta S(T) =  \int \Delta \sigma_{1}(\omega)d\omega$
 where the integral is taken for the relevant ranges 153~meV 
 $\sim$ 186~meV 
for I and 186~meV $\sim$ 260~meV 
 for II, respectively. Peak I grows reorganizing the spinon spectral
 weight which is transferred from low energies to high energies. In
 addition the whole magnetic line shape gains weight as the temperature is lowered.
This is due to transfer of spectral weight from the phonon region to
the magnetic region as the AF ordering builds up and the excitation of
antialigned spins become more likely. Notice that these two processes
keep the total sum rule practically constant
 [c.f. Fig.~\ref{fig:fig5}(c)]. On passing  $T_{N1}$ we notice
 that an anomaly appears on the temperature
 dependence of peak II when the spiral develops at $T_{N2}$,
which may be related to the mutiferroic effect.

The temperature dependence of peak I shows that it requires long-range
order where spin waves became well defined propagating excitations.
The dipole operator produces two spin waves
in nearest neighbor sites which interact strongly\cite{lor95,lor95b}.
A rough estimate of the position of the peak
can be obtain from the energetic cost to flip two neighboring spin,
along the AF chain, in
the Ising limit. Taking into account only  $J$ along the chains
one expects the peak
to appear at $\omega_{ph}+J= 0.159$~eV
in good agrement with the observation. Clearly the smaller
interactions which drive the 3D AF order can not be neglected and
at first sight they would spoil the agreement. Their
effect would be to increase (decrease) the estimate providing they are
 non-frustrating  (frustrating) with the non-frustrating effect
 expecting to dominate. To some extent this is already
 taken into account by the renormalization of $J$ however some
 hardening is still expected. On the other hand  more realistic interacting
 spin wave theory (ISWT) computations  give generally a smaller result
 than the Ising estimate so a cancellation of errors may explain the
surprisingly good agreement of the simple estimate $\omega_{ph}+J$. For example
 in 2D one obtains $\Delta E_{ISWT}=2.7J$ while   $\Delta
 E_{Ising}=3J$. A precise ISWT computation in the present case is not
 possible because the structure is much more complicated and the
 interchain interactions are not precisely known.

In the ordered temperature range
 $\Delta \sigma_{1}(\omega)$ exhibits two weak sideband peaks at higher
frequencies  as indicated by
the arrows. Their peak energies, 0.276eV  and 0.334eV are close to
$\omega_{ph}+2J=0.216$eV and $\omega_{ph}+3J=0.316$eV  consistent
with  multi-spin-wave excitations as observed in 2D cuprates.
Also the position of peak II has an anomaly at the
ordering temperature and hardens below $T_{N1}$.
All this shows that the reorganization of the spinon spectral weight
due to the emergence of the spin waves occurs also at high
energies. Spinons and spin-waves  mix but conserve to a
surprising extent their own character.

%
%

The value found for  $J\sim 95-100$~meV is larger than
a previous ``rough'' neutron scattering estimate\cite{yan89b} (67$\pm$20~meV), a
susceptibility measurement\cite{shi03} (73$\pm$ 3~meV) and
self-interaction-free LDA (52~meV)\cite{fil05}
but smaller than a recent LDA study
(126~meV or 118~meV depending on detail of the method)\cite{roc09}.
Notice also that our measurements weight more heavily the high energy part of the spectrum which
may be affected differently by the smaller interactions than the previous
experiments.

To conclude we have shown that CuO can be described at high
temperatures as a 1D quantum antiferromagnet with fermionic
excitations. As the  temperature is lowered a dimensional crossover
occurs and the system becomes ordered with well defined propagating
spin-waves but with a strong remnant of the spinon band. A similar
phenomena has been seen in KCuF$_3$ using neutron scattering where the
spin waves appeared as a very subtle reorganization of the low energy spinon
spectral weight\cite{lak05}. Here the effect is much more dramatic challenging
theories of the crossover.

 The ordered
spectrum is qualitatively similar to the one of AF cuprates parent
of the high-$T_c$ superconductors. There the spectrum was
interpreted as the superposition of two-magnon propagating spin wave
excitations plus a prominent high energy band due to incoherent
multimagnon processes, for which no well establish theory exist.
Comparison with CuO suggest that a spinon interpretation is
reasonable\cite{cho01}. Furthermore, the relative weight of the
two propagating spin wave processes respect to the higher energy excitations
decreases as one approaches spin liquid states changing the spin or
the dimensionality.  For example in spin-1 2D La$_2$NiO$_4$ the spectrum is
exhausted by the two-magnon peak\cite{per98}, in 2D cuprates  43\%
of the spectral weight is in the two-magnon peak\cite{lor99}, here it
amounts to only $\sim 10\%$ and in Sr$_2$CuO$_3$\cite{lor97} it is
zero. This progression gives credit to theories which argue that 2D
cuprates are close to a spin liquid phase.

This work was supported by the Frontier Research Facility Program at
the University of Seoul. Y. S. and T. K. were supported by KAKENHI
(20674005 and 20001004), Japan. J. L was supported by MIUR PRIN
2007FW3MJX003 and partially by NSF grant PHY05-51164 at KITP. J.L
thanks KITP-UCSB for hospitality under the program ``The Physics of
Higher Temperature Superconductivity''.


\end{document}